\documentclass{revtex4}


\usepackage{amsmath}

\begin{document}

\title{Clausius relation for active particles: \\ what can we learn from fluctuations?}

\author{Andrea Puglisi}
\affiliation{CNR-ISC and Dipartimento di Fisica, Sapienza Universit\`a di Roma, p.le A. Moro 2, 00185 Roma, Italy; andrea.puglisi@roma1.infn.it}

\author{Umberto Marini Bettolo Marconi}
\affiliation{Scuola di Scienze e Tecnologie, Universit\`a di Camerino, Via Madonna delle Carceri, 62032, Camerino, INFN Perugia, Italy; umberto.marinibettolo@unicam.it}






\begin{abstract}
  Many kinds of active particles, such as bacteria or active
  colloids, move in a thermostatted fluid by means of
  self-propulsion. Energy injected by such a non-equilibrium force is
  eventually dissipated as heat in the thermostat. Since thermal
  fluctuations are much faster and weaker than self-propulsion forces,
  they are often neglected, blurring the identification of dissipated
  heat in theoretical models. For the same reason, some freedom - or
  arbitrariness - appears when defining entropy production. Recently
  three different recipes to define heat and entropy production have
  been proposed for the same model where the role of self-propulsion
  is played by a Gaussian coloured noise. Here we compare and discuss the relation
  between such proposals and their physical meaning. One of these
  proposals takes into account the heat exchanged with a
  non-equilibrium active bath: such an ``active heat'' satisfies the
  original Clausius relation and can be experimentally verified.
\end{abstract}



\maketitle

\section{Introduction}

Active particle systems have recently attracted the increasing
interest of scientists of different disciplines since they sit at the
intersection of biology, chemistry and physics~\cite{ram10,marchetti2013}.
A central feature of these materials is that its elementary
constituents convert energy from the environment via metabolic or
chemical reactions into direct motion but also dissipate energy
producing heat by friction in order to move inside a surrounding
solvent~\cite{BLLRVV16}.  Therefore, the complex behavior of active
particles can only be described by the tools of non equilibrium
statistical physics, such as kinetic theory, statistical mechanics of
non-equilibrium processes and stochastic
thermodynamics~\cite{cates2012diffusive}.  A fascinating question,
which naturally comes to our mind, is how thermodynamics shapes
biological functions in living organisms~\cite{england} such as
motility and self-propulsion and in particular which is the entropy
production associated with their non equilibrium steady
states~\cite{DEGM}.  This question requires a notion of heat generated
by self-propulsion and dissipated in the thermostatted solvent.

Within careful calorimetric experiments, one is able to measure the
heat dissipated into the solvent by a microbial
colony~\cite{calor}. Such total heat depends upon many biological
functions which are not included in active models, but one could devise
smart experiments (e.g. by varying motility without changing other
functions) in order to assess the fraction of heat strictly generated
by self-propulsion. Observing the associated fluctuations is perhaps a
much harder task, if not impossible. However, before encountering
experimental limitations, one finds limits in the theory. 

At our mesoscopic level, the definition of heat has to be framed
within stochastic
thermodynamics~\cite{LS99,seifert05,BPRV08,sekimoto}. A problem,
however, arises when thermal fluctuations are discarded: such an
approximation is adopted in many models of active particles, since
temperature is negligible with respect to the energy associated with
both self-propulsion and external forces. In active models, some noise
is retained to describe the non-deterministic nature of the
self-propulsion force, but it usually acts on time-scales and energy
scales much larger than molecular agitation of the solvent. This
``athermal'' nature of active particle models is similar to that in
granular models~\cite{PVBTW05} or in models of macroscopic
friction~\cite{BTC11}. While it is very useful - sometimes even
inevitable - for the purpose of analytic calculations or numerical
simulations, it leads to a mismatch between entropy production and
heat~\cite{CP15}. Basically, the relation between total entropy
production and dissipated heat loses its similitude with the original
Clausius form and involves additional terms. For this reason, it does
not provide a clear constraint on heat divided by temperature, as it
occurs for the Clausius relation in macroscopic thermodynamics. Such a
problem has already been noticed in some models of active
particles~\cite{GC13}, and in systems with
feedback~\cite{KQ04,MR12,MR14}.

Recently it has been shown that the above fallacy is bypassed in a
model of active particles where Gaussian colored noise plays the role
of self-propulsion~\cite{MBGL15,BPM17}. Even if thermal fluctuations
are neglected, a notion of ``coarse-grained heat'' can be introduced,
together with a spatial-dependent effective temperature, such that the
original Clausius relation is fully recovered. The crucial point here
is that both such a ``coarse-grained heat'' and the effective
temperature can be measured in experiments and therefore a test of
this active Clausius relation can be attempted. In the last year other
two proposals have appeared in the literature~\cite{FNCTVW16,mandal},
devoted to define entropy production and heat in the same identical
model. The purpose of the present paper is to discuss the connection
between those different Clausius relations. Apart from this
comparison, two novelties are present here with respect
to~\cite{BPM17}: 1) a single trajectory level of description is
adopted, while in~\cite{BPM17} an ensemble average had been
considered; 2) a generalization of~\cite{BPM17} to more than one
dimension and interacting particles is presented. In order to simplify
the discussion, the main discussion is focused on the 1d case with a
time-independent potential, which is sufficient to show the main
difference between the definitions of heat and entropy production in
the three works considered~\cite{FNCTVW16,BPM17,mandal}. The
generalisation of the proposal in~\cite{BPM17} to multi-dimensional
cases with a time-dependent potential is also discussed at the end.

In Section 2 we review the basic facts of Clausius relation in
macroscopic thermodynamics and in stochastic (or mesoscopic)
thermodynamics. In Section 3, we introduce the active model with
Gaussian colored noise with thermal fluctuations, where a
``microscopic Clausius relation'' is trivially satisfied, and then
show what happens at the coarse-grained level, when inertia and
thermal fluctuations are neglected. The three recipes appeared in~\cite{FNCTVW16,BPM17,mandal} to connect
entropy production, dissipated heat and temperature are reviewed and
compared.

\section{Heat and entropy production: from macroscopic to stochastic thermodynamics}

Here we revise a few elementary facts of thermodynamics, at the
macroscopic level and at the mesoscopic one. The macroscopic level is
the one presented in thermodynamic textbooks, where there are no
fluctuations: we denote quantities at this level with capital
letters. The mesoscopic level is the topic of an intense research
exploded roughly in the last two decades, and is dominated by
fluctuations: we denote quantities at this level with small
letters. Averaging out the fluctuations
of the mesoscopic level (an operation which - in general - is
automatically obtained in the limit of a very large number of
constituents) brings back the results of the macroscopic one. Through
the whole paper, we set the Boltzmann constant $k_B=1$.

\subsection{Macroscopic level}

For a large system one may measure heat, for instance with a
calorimeter, and call it $\delta Q$: we consider it positive when going
from the bath to the system. The second principle of thermodynamics guarantees that in a
 transformation where entropy $S$ changes by a quantity $dS$ there is a non-negative entropy production~\footnote{We use a shorthand
  notation where - for instance $\delta \Sigma$ and $\delta Q$ are
  non-exact differentials, while $dS$ is an exact differential.}
\begin{equation} \label{secp}
\delta \Sigma = dS-\frac{\delta Q}{T} \ge 0.
\end{equation}
In a ``quasi-static'' transformation the equal sign holds, i.e. $\delta \Sigma=0$.

In the absence of a direct way of measuring the entropy of a system
(e.g. if quasi-static transformations are not available), other
relations derived from Eq.~\eqref{secp} and involving measurable
observables are useful. For instance from Eq.~\eqref{secp} it follows
\begin{equation}
dS \ge \frac{\delta Q}{T}
\end{equation}
and therefore the existence of a ``minimum work'' that can be extracted, also called difference of free energy $dF$:
\begin{equation}
T dS \ge dE - \delta W \rightarrow \delta W \ge dE-TdS=dF.
\end{equation}
Another consequence that is derived way from Eq.~\eqref{secp} is to consider a cyclical transformation ($dS=0$), where it implies
  \begin{equation}
\oint \frac{\delta Q(t)}{T(t)} \le 0,
  \end{equation}
which is the celebrated Clausius relation. This can be tested in
experiments and is the founding principle of the theory of heat
engines, efficiency, etc.

It is important to underline that if - hypothetically - the total entropy production was something different from $dS-\delta Q/T$, i.e. if
\begin{equation} \label{anom}
\delta \Sigma = dS-\frac{\delta Q}{T} + \Sigma_{an} \ge 0,
\end{equation}
then all the above relations, including the Clausius relation, would
not hold anymore due to the presence of an "anomalous'" entropy
production term, $\Sigma_{an}$. However it is quite difficult to
imagine Eq.~\eqref{anom} in macroscopic thermodynamics, since the very
definition of macroscopic entropy production is the difference between
$dS$ and $\delta Q/T$~\cite{DEGM}. On the contrary, equations similar
to~\eqref{anom} have appeared in the literature in a stochastic
thermodynamic treatment of systems with feedback and model of self-propelled particles~\cite{KQ04,MR12,GC13,MR14,mandal}.

\subsection{Mesoscopic level}
\label{mes}

When a small system is considered, a stochastic description is
necessary in order to incorporate fluctuations. At thermodynamic
equilibrium the stochastic evolution must be consistent with
micro-reversibility. More precisely, the couple ``system plus
thermostat'' describes all degrees of freedom of the world and
therefore it has to satisfy an exact symmetry under time-reversal:
when the heat bath is replaced by an effective stochastic bath force,
time-reversal is mapped into the equivalence of probabilities of a
trajectory and its time-reversal, which coincides with detailed
balance if the process is Markovian.


When an external, non-conservative, force is applied to the system,
one may expect that the stochastic bath force is not changed (for
instance if the bath is very large and is weakly affected by the
external force). This amounts to say that the non-equilibrium model
contains the sum of two forces which both concur to change the energy of the
system: the external force does work, the bath force brings
heat~\cite{sekimoto}. A notion of entropy production rate $\sigma$ of
a trajectory $\omega(t)$, for Markovian stochastic systems, has been
introduced in~\cite{LS99} and revisited in~\cite{seifert05}. It can be
summarized as
\begin{align} \label{ls}
\int_{0}^t \delta \sigma(t') &= \ln \frac{\textrm{prob}[\{\omega(t')\}_{0}^t]}{\textrm{prob}[\{\overline{\omega}(t-t')\}_{0}^t]} = \ln\frac{p[\omega(0)]}{p[\overline{\omega(t)}]}+ \ln \frac{\textrm{prob}[\{\omega(t')\}_{0}^t|\omega(0)]}{\textrm{prob}[\{\overline{\omega}(t-t')\}_{0}^t|\overline{\omega}(t)]}  = \\ &= \int_0^t ds + \int_0^t \delta s_m,
\end{align}
where $\overline{\omega}$ is the time-reversal of the phase-space
variables (typically positions are unchanged and velocities are
reflected), $s(t)=-\ln p[\omega(t)]$ is the microscopic Gibbs entropy
in the point $\omega(t)$ in phase space and $\delta s_m$ is the
so-called entropy production of the surrounding
medium~\cite{seifert05}. In the rest of the paper we consider, for simplicity, the infinitesimal version of Eq.~\eqref{ls}, i.e.
$\delta\sigma = ds + \delta s_m$.
An average over noise realizations and initial conditions is expected
to give back the macroscopic quantities, i.e. $\delta\Sigma = \langle \delta \sigma
\rangle$ and $dS = \langle ds \rangle$, such that Eq.~\eqref{secp}
implies $\langle \delta s_m \rangle = -\delta Q/T$.  Indeed in many
models at constant temperature, one has $\delta s_m = -\delta q/T$ with
$\delta q$ the heat injected by the bath force, which satisfies
$\langle \delta q \rangle = \delta Q$. The total entropy production
$\int_0^t \delta\sigma(t')$ satisfies the Fluctuation-Relation at any
time $t>0$ and this guarantees that  $\delta \Sigma=\langle \delta \sigma \rangle$ is
non-negative~\cite{LS99}. In a stationary state $\langle \delta s_m \rangle \ge 0$
then follows.

As a useful example, let us consider the evolution of a colloidal
particle of mass $m$, position and velocity $x(t),u(t)$ in one
dimension, under the action of an external potential $\phi(x)$ and of
a non-conservative external force $f_{nc}(t)$. 
\begin{subequations} \label{generic}
\begin{align} 
dx(t) &= u(t)dt\\
  m du(t)&=-\gamma u(t)dt +\sqrt{2\gamma T}dW(t)-\phi'[x(t)]dt+f_{nc}(t)dt,
\end{align}
\end{subequations}
with $dW(t)$ the Wiener infinitesimal increment (with variance $dt$). Defining
energy as $e=mu^2/2 + \phi(x)$, it is easy to see that heat (going
from the bath into the system) reads
\begin{equation}
\delta q = de - \delta w = u \circ [-\gamma u dt+\sqrt{2\gamma T}dW(t)]
\end{equation}
where we have defined the work $\delta w=u f_{nc} dt$, and $\circ$ denotes products which must be integrated according to the Stratonovich rule.

For this model, it is possible to compute the conditional probability
appearing in~\eqref{ls} and therefore compute $\delta s_m$. The result
depends upon the parity of $f_{nc}$ under time-reversal~\cite{CG07}.
In simple cases, for instance when
magnetic fields are not involved~\cite{pradhan10}, such a force is
assigned even parity under time-reversal. In this case one gets (see Appendix)
\begin{equation} \label{seifert}
\delta \sigma = ds - \frac{u \circ [-\gamma u dt +\sqrt{2\gamma T}dW]}{T} =  ds -\frac{\delta q}{T},
\end{equation}
which is $\ge 0$ on average, leading to the usual Clausius relation.

On the contrary if $f_{nc}$ is odd, for instance if the
coarse-graining has delivered a force which is proportional to odd powers of the velocity
of external bodies, or if magnetic fields are involved~\cite{pradhan10}, the relation~\eqref{seifert} does not hold anymore.
In such cases, things seem to improve when the so-called {\em
  conjugated} dynamics is considered, by changing the sign of odd
external non-conservative forces when computing the probability of
inverse paths appearing in the denominator of
Eq.~\eqref{ls}~\cite{jarzy2006,CG07,ford2012,MR14}: basically this
amounts to change the parity of the force and get back the result in
Eq.~\eqref{seifert}.  The problem of such an artificial prescription,
however, is that the conjugated dynamics cannot be realized in
experiments and therefore an empirical evaluation (i.e. without a
detailed knowledge of the equation of motions) of the conjugated
probability is not available, neither it is possible to experimentally
observe the associated fluctuation relation.

\section{Active particles: the coarse-grained heat and Clausius relation}

The analogy between stochastic and macroscopic thermodynamics,
Eq.~\eqref{seifert}, rests upon two main ingredients: 1) the heat bath
must be modeled as a stochastic force which - if non-conservative
forces are removed - satisfies detailed balance with respect to the
equilibrium probability distribution ($\delta \sigma \equiv 0$) and 2) the non-conservative
forces are even under time-reversal, a fact which is expected to be
realized when the microscopic forces are not velocity-dependent
(e.g. there are no Lorentz forces) and the coarse-graining does not
change or mix their parity. Many models of active particles
abandon such basic facts (in particular detailed balance~\cite{cates2012diffusive}), with the
aim of describing the relevant variables (such as positions or
orientations of the micro-swimmers) which evolve on scales much slower
than those affected by thermal agitation. An interesting example of
model of active swimmers where this procedure can be analyzed is one where self-propulsion takes the form
of an Ornstein-Uhlenbeck process: its non-zero correlation time
represents the persistence of motion due to activity.

Here we introduce the model at a space-time scale fine enough to
describe the real velocity $u$ of the particle and thermal
fluctuations:
\begin{subequations}\label{underd}
\begin{align} 
dx(t)&=u(t)dt \\
mdu(t)&=-\gamma u(t)dt + \sqrt{2\gamma T_b}dW(t)+ f_a(t)dt-\phi'[x(t)]dt,
\end{align}
\end{subequations}
where $T_b$ is the environmental (solvent) temperature and the active force satisfies
\begin{equation}
df_a(t) = -\frac{f_a(t)}{\tau}dt + \frac{\gamma\sqrt{2D_a}}{\tau}dW_2(t),
\end{equation}
with $dW_2$ another (independent) Wiener increment with variance
$dt$. Here we consider for simplicity the $1$-particle case in one
dimension, with a potential $\phi(x)$ which does not depend upon
time. Later we generalize some of the results to many interacting
particles and with a time-dependent potential.

Note that, when $\phi=0$, $\langle x^2 \rangle \sim 2(D_a+T_b/ \gamma)
t$ for large times. Based upon such a bare diffusivity, the ``active
bath temperature'' $T_a=\gamma D_a$ is usually
defined~\footnote{In~\cite{BPM17} a ``mass-less'' active temperature
  $T_\tau=D_a/\tau$ was defined. In this paper we show that it is not
  necessary, if an ``effective mass'' $\mu=\gamma \tau$ is used, as
  in~\cite{mandal}.}. Of course there is no thermostat at temperature
$T_a$, such a temperature is only useful to define a relevant energy
scale.

Since active micro-swimmers are usually dispersed in viscous liquids,
it is much more common to find the overdamped version of the
model~\cite{MBGL15}, which describes the position of the
particle on a time-scale slower than the relaxation time due to
inertia:
\begin{equation} \label{overd}
dx(t)=\frac{\sqrt{2\gamma T_b}dW(t)+f_a(t)dt-\phi'[x(t)]dt}{\gamma}.
\end{equation}

\subsection{Heat dissipation into the solvent}

Interpreting $f_a$ as an external force derived - through the
coarse-graining of the full microscopic dynamics - from forces which
do not depend upon velocities, it is reasonable to consider it
even. According to the recipe of stochastic thermodynamics discussed
above, Eq.~\eqref{ls} applied to Eq.~\eqref{underd} or
Eq.~\eqref{overd}, see Appendix, one gets Eq.~\eqref{seifert}, with
\begin{equation}
\delta q_b = u \circ [-\gamma u dt+\sqrt{2\gamma T_b}dW(t)],
\end{equation}
which is the heat absorbed from the reservoir, satisfying in the steady state the Clausius relation at constant temperature, i.e.
\begin{equation}
\delta Q_b = \langle \delta q_b\rangle \le 0.
\end{equation}
The interpretation is obvious, the active force $f_a(t)$ acts as an external
non-conservative force and transfers energy in the system which is
dissipated into the bath. This can be measured by ordinary calorimetry
in the solvent~\cite{S16}. As discussed above, such a measurement is
in principle very difficult in experiments with living micro-swimmers,
since released heat is affected by many other non-equilibrium
biological functions. A promising direction could be the use of
artificial active particles~\cite{BLLRVV16}.

\subsection{Removing the solvent from the description}

Since $T_b$ is orders of magnitude smaller than active temperatures,
it is very useful - also for computational purposes - to remove it from Eq.~\eqref{overd}, keeping only
\begin{equation} \label{simplest}
\dot{x}=\frac{f_a(t)-\phi'(x)}{\gamma}.
\end{equation}
At this point an important ingredient of the bath force (its noise)
has disappeared and the basic recipe of stochastic thermodynamics
cannot be applied straightforwardly. Still, it is useful to find a measure of
``distance from equilibrium'' and relate it to parameters and
observable quantities.  Considering that $f_a$ is random, one is
tempted to consider $-\gamma \dot{x}+f_a(t)$ as an effective bath and
define a heat as $\dot{x} \circ [-\gamma \dot{x}+f_a(t)]$. However,
the random force $f_a(t)$ is non-Markovian and therefore the standard recipe of stochastic
thermodynamics brings in complications~\cite{ZBCK05,SS07,CPV12}.

The simplest way to get rid of the non-Markovian character of the noise is to time-derive Eq.~\eqref{simplest}, obtaining
\begin{subequations}
\label{int}
\begin{align}
dx(t)&=u(t)dt\\
\label{int1} \mu du(t)&=-\gamma u(t)dt +\sqrt{2 \gamma T_a}dW(t)-\phi'[x(t)]dt-\tau \phi''[x(t)] u(t)dt = \\ &= -\gamma\Gamma(x) u(t)dt+\gamma\sqrt{2 D_a}dW(t)-\phi'[x(t)]dt \label{int2}
\end{align}
\end{subequations}
where we have introduced the effective mass $\mu=\gamma\tau$ and the space-dependent viscosity correction $\Gamma(x)=1+\frac{\tau}{\gamma}\phi''(x)$.

As highlighted by the two versions in Eqs.~\eqref{int1}-\eqref{int2},
the evolution of the effective velocity $u$ is affected by the
conservative force $-\phi'(x)$ and by an additional force that can be
interpreted in two different ways: 1) an equilibrium bath at
temperature $T_a$ plus a non conservative force $f_{nc}=-\tau\phi''(x)
u$ which is {\em odd} under time-reversal, or 2) a non-equilibrium
bath with space-dependent viscosity modulated according to the
function $\Gamma(x)$. In the next two subsections, we see the
consequences of such different interpretations, which change both the
definition of entropy production as well as of heat.

\subsubsection{Equilibrium bath with a non-conservative force: conjugated entropy production}

This interpretation is considered in~\cite{mandal}. The authors
propose to define heat as the energy injected by the force $-\gamma u dt + \gamma \sqrt{2D_a}dW$, as if it were an
equilibrium bath
\begin{equation}
\delta q_1 = u \circ (-\gamma u dt + \gamma \sqrt{2 \gamma T_a}dW).
\end{equation}
To derive the entropy production, the authors consider the
formula~\eqref{ls} with the probability of the time-reversed path
(which appears in the denominator) computed according to a dynamics
where the force $f_{nc}(t)$ is replaced by $-f_{nc}(t)$, as discussed
at the end of Sec.~\ref{mes}. This idea is justified by the authors by
showing that such a change of sign is necessary in order to make
invariant under time-reversal the dynamics without the bath. However
such an argument is not really compelling. The terms $-\gamma u dt +
\gamma \sqrt{2\gamma T_a}dW$ do not correspond to any well-defined
part of the physical system which could be identified as an
equilibrium bath: the first term is the viscous damping due to the
solvent, the second term is the fluctuating part of the derivative of
the self-propulsion. It is a mathematical coincidence that together
they form a Ornstein-Uhlenbeck process of the same form of equilibrium
bath forces. In our opinion, it is quite arbitrary detaching them from
Eq.~\eqref{int} and there is no reason why the rest of the equation
(once those terms are removed) should satisfy the invariance under
time-reversal.

According to the ``conjugated'' prescription, one gets for the case of a single particle considered here (see Appendix)
\begin{equation} \label{mand}
\delta \Sigma = ds-\frac{\delta q_1}{T_a} + \frac{\tau^2}{2 T_a } (du)^2 \phi'',
\end{equation}
where $(du)^2 \approx 2 T_a dt/(\gamma \tau^2)$.
The average can be written as
\begin{equation} \label{notcla}
\delta \Sigma = dS-\frac{\delta Q_1}{T_a}+\frac{dt}{\gamma}\langle \phi'' \rangle \ge 0,
\end{equation}
with $\delta Q_1=\langle \delta q_1 \rangle $.

A peculiarity of this recipe is that it gives a non-zero average
entropy production also for the harmonic case $\phi(x) \sim
x^2$. Since in the harmonic case Eq.~\eqref{int} satisfies detailed
balance, it is unclear if such a peculiarity is an advantage or
not. Moreover, as already discussed, entropy production computed with
the conjugated reversed dynamics is not accessible in
experiments. Most importantly, in our opinion Eq.~\eqref{notcla}
hardly deserves the name ``Clausius relation'', as it does not give the same important information about the sign of the average heat.

\subsubsection{Equilibrium bath with a non-conservative force: standard entropy production}

In~\cite{FNCTVW16} the authors consider formula~\eqref{ls} (without
conjugation for the reversed dynamics) applied to the dynamics in Eq.~\eqref{int2}. However all terms giving exact
deterministic differentials are thrown away, leading to an approximate
formula (see Appendix)
\begin{equation} \label{fred}
\delta \Sigma \approx \frac{\tau^2 u \phi'' \circ du}{T_a}.
\end{equation}
In the steady state the neglected terms have zero average,
and indeed only the average formula is reported
in~\cite{FNCTVW16}. Fluctuations and large deviations functions, however, may
keep the memory of those terms~\cite{visco,PRV06,BGGZ}.

Another difficulty of formula~\eqref{fred} is its connection with
heat. In the end of their paper, the authors manage to show that
Eq.~\eqref{simplest} can be mapped exactly into a generalized Langevin
equation with memory. This equation can be broken into a viscoelastic
bath at equilibrium at temperature $T$ plus a non-conservative
force. Within such a description, the average of the entropy
production rate in Eq.~\eqref{fred} can be written as $\mathcal{J}/T$,
where $\mathcal{J}$ is the heat flux dissipated into the bath. A
simple formula for such a ``viscoelastic'' heat or its - local or
global - average is not given in~\cite{FNCTVW16}. Most importantly,
some terms of fluctuations of entropy production are neglected which
could be relevant for large deviation functions and the validity of the Fluctuation Relation~\cite{visco,PRV06,BGGZ}.

\subsubsection{Non-equilibrium bath}

If the standard recipe of stochastic thermodynamics, Eq.~\eqref{ls}, is used without neglecting any term, one gets (see Appendix)
\begin{equation} \label{newcla}
\delta \sigma = ds - \frac{\delta q_2}{\theta(x)},
\end{equation}
with the ``active bath heat'' defined as
\begin{equation} \label{newhe}
\delta q_2 = u \circ d f_{ab},
\end{equation}
the ``active bath force'' as
\begin{align}
  df_{ab}(t) &= -\gamma\Gamma[x(t)] u(t) dt+ \gamma\sqrt{2D_a} dW(t)  =\\ 
        & = -\gamma\Gamma[x(t)] u(t) dt+ \sqrt{2  \gamma \Gamma[x(t)] \theta(x)} dW(t)
\end{align}
and the ``local active temperature'' $\theta(x) = T_a/\Gamma(x)$. This
interpretation is supported by the observation that a local
Maxwellian with temperature $\theta(x)$ is an approximate solution for the local velocity
distribution, with ``small'' violations of detailed balance,
see~\cite{BPM17} for details. 

Averaging Eq.~\eqref{newcla} at fixed position $x$, one gets
\begin{equation}
\delta \Sigma = dS- \frac{\langle \delta q_2(x) \rangle}{\theta(x)} \ge 0,
\end{equation}
which in the steady state ($dS=0$) is identical to the Clausius
relation. In~\cite{BPM17}, the ensemble average of the above relation
over the steady state probability distribution $p(x,u)$ has been considered. Interestingly, the local average $\dot{\tilde q}(x)$ of the dissipated heat flux reads
\begin{equation} \label{tildeq}
\dot{\tilde q}(x) = \gamma\Gamma(x)\left[\frac{\theta(x)}{\mu} n(x)-\int dv u^2 p(x,u)\right ],
\end{equation}
where $n(x)=\int dv p(x,u)$. This is an additional argument in favour
of the simplicity and consistency of the picture discussed in the
present section: the ``active heat'' is exactly proportional to the
difference between the local active temperature $\theta(x)$ and the
empirical temperature $\langle u^2 \rangle_x$. The empirical
temperature is ``attracted'' by the local active temperature but the non-uniformity of such a temperature prevents full relaxation: the
mismatch is a source of flowing heat. Eq.~\eqref{tildeq} shows a straightforward way to measure such ``active heat''.

We note that when the potential does not depend upon time, as in all
our equations up to this point, the active heat $\delta q_2$ has zero
average. Nevertheless, the average entropy production
$\delta \Sigma$ has non-zero average, apart from the harmonic case
$\phi(x) \sim x^2$ which is a special case where $\theta(x)$ is
uniform~\cite{BPM17}. 

When more particles are involved, a (local and time-dependent)
diagonalisation procedure can always set back the problem in the case
of a single particle. The multi-particles and multi-dimensional
version of Eq.~\eqref{int} reads
\begin{equation}
\mu d u_i = -\gamma \Gamma_{ij}(\mathbf{r})u_j dt -\partial_i \phi(\mathbf{r}) dt + \gamma \sqrt{2D_a} dW_i,
\end{equation}
with $\Gamma_{ij}=\delta_{ij}+\frac{\tau}{\gamma}\partial_j \partial_i
\phi$ and indexes running over all particles and all Cartesian
components and the Einstein summation convention is assumed. The potential $\phi$ includes both external and internal
forces. Since the matrix $\Gamma_{ij}(\mathbf{r})$ is symmetric, an
orthogonal matrix $P_{ij}(\mathbf{r})$ always exists such that
$P\Gamma P^T=D$ with $D_{ij}(\mathbf{r})=\lambda_i(\mathbf{r})
\delta_{ij}$. By defining the rotated coordinates ${\mathbf R}=P
      {\mathbf r}$ and velocities ${\mathbf U}=P
      {\mathbf u}$, and recalling that the gradient rotates as a
      vector and the rotation of the vector of independent white
      noises gives again a vector of independent white noises, it is
      straightforward to get the formula:
      \begin{equation}
\mu dU_i = -\gamma \lambda_i({\mathbf R}) U_i dt - \partial_{R_i} \phi + \gamma \sqrt{2 D_a} dW_i.
        \end{equation}
      Computation of the entropy production leads, therefore, to
      \begin{equation}
      \label{thetai}
\delta \sigma(t) = ds(t) - \sum_i \frac{\delta q_{2,i}(t)}{\theta_i[{\mathbf{R}}(t)]},
      \end{equation}
      with $\theta_i({\mathbf R})=T_a/\lambda_i(\mathbf{R})$ the $i$-th component of the local active temperature and
\begin{equation}
\delta q_{2,i} = U_i \circ [-\gamma \lambda_i({\mathbf R}) U_i dt + \gamma \sqrt{2 D_a}dW].
\end{equation}
Notice that  eq.\eqref{thetai} generalizes the Clausius relation to a system with different temperatures $\theta_i$.
      
As an example, in the case of  an active particle moving in a plane a subject to a central potential $\phi(r)=\phi({\mathbf r})$,
we have the following Cartesian representation of the matrix
$D_{ij}(r)=D_r(r) \hat r_i\hat r_j +D_t(r)(\delta_{1j}-\hat r_i\hat r_j )$ with $D_r(r) =1+\frac{\tau}{\gamma}  \phi''(r)$
and $D_t(r) =1+\frac{\tau}{\gamma}  \phi'(r)/r$. The two temperatures are $\theta_r(r)=1/D_r(r)$ and $\theta_t(r)=1/D_t(r)$.

\subsection{Time-dependent potential}

When an external transformation is considered,
i.e. a time-dependent potential $\phi(x,t)$ is taken into account,
Eq.~\eqref{int2} is replaced with
\begin{equation}
\mu du(t)= -\gamma\Gamma(x) u(t)dt+\gamma\sqrt{2 D_a}dW(t)-\partial_x\phi[x(t),t]dt - \tau\partial_t \partial_x \phi[x(t),t]dt \label{int2bis}
\end{equation}

Also, in this case, we get (see Appendix) the validity of the
mesoscopic Clausius relation Eq.~\eqref{newcla} with the active heat
Eq.~\eqref{newhe}. Time-dependent potentials are at the
basis, for instance of realizations of heat engines~\cite{wij17}.

\subsubsection{Very slow transformations}

Imagine a very slow transformation from a $\phi(x,t_1)$ to a new
$\phi(x,t_2)$: this means transforming the original non-equilibrium
steady state (``NESS'', at $t<t_1$) to a new non-equilibrium steady
state (for $t\gg t_2$). As discussed above, in the initial and final
NESS there is heat going steadily to the bath, even without the
transformation. Therefore for very slow transformations $\Delta Q \to
- \infty$ and the Clausius relation becomes useless. For this reason Oono-Paniconi~\cite{PO98}, then
Hatano-Sasa~\cite{HS01}, Bertini et al.~\cite{bertini14} and
Maes~\cite{maes14} have found expressions for the so-called ``excess
heat'', i.e. heat which is released for the sole purpose of the
transformation: this heat is obtained removing the ``housekeeping
heat'' (necessary for the steady states) from the total $\Delta
Q$.
All the mentioned
proposals have been given for overdamped systems, where certain
symmetries are more clear but also less general. Active particles have
some kind of inertia or persistence which cannot be disregarded and
therefore do not comply with such an assumption. It would be
interesting to see the above simple ideas applied to the model in Eq.~\eqref{simplest}  with a
slow transformation of the potential.

\section{Conclusions}

In this paper, we have reviewed and compared three different
prescriptions to extend the Clausius formula to active systems,
i.e. to particles able to self-propel by means of metabolic processes
or chemical reactions and to dissipate energy by a frictional
mechanism. Those relations between the heat dissipation, entropy and
work appear at the mesoscopic level, where fluctuations are taken into
account by means of a stochastic description, but the
contribution to these fluctuations coming from the molecular bath is
neglected, leaving a certain freedom in defining heat and entropy production.

While all three methods are admissible and do not contradict any
general principle, our point of view gives indications that only one
of these prescriptions can be considered as a stochastic version of the
original Clausius heat theorem, that is Eq.~\eqref{newcla} with
``active heat'' defined in~\eqref{newhe}. The stochastic "active"
version of Clausius formula we have derived coincides with the one
recently presented by using ensemble averaged quantities~\cite{BPM17}.

\acknowledgments{We warmly akcnowledge useful discussions and communications with L. Cerino, A. Sarracino and F. van Wijland. }

\appendix

\section{Entropy production}

We consider here a generalization of the dynamics in
Eq.~\eqref{generic} with $f_{nc}(x,u,t)$ representing any kind of
time-dependent term: it can be external or internal (that is
function also of system's degrees of freedom), odd or even under
time-reversal, and we consider both the normal or the conjugated dynamics
for the probability of the time-reversed path. In particular we
assume $dx=u dt$ and
\begin{align}
m du = d\alpha(x,u,t) = -\gamma u(t) dt + \sqrt{2\gamma T}dW(t)-\phi'[x(t)]dt+f_{nc}(x,u,t)dt \\
m du = d\alpha^*(x,u,t) = -\gamma u(t) dt + \sqrt{2\gamma
  T}dW(t) -\phi'[x(t)]dt+\overline{f}_{nc}(x,u,t)dt 
\end{align}
to generate the dynamics of the forward and reversed paths,
respectively. The Wiener increments $dW(t)$ have variance $dt$. When
the standard entropy production is computed, the dynamics is the same,
i.e. $\overline{f}_{nc}(x,u,t)=f_{nc}(x,u,t)$. On the contrary for the
conjugated entropy production
$\overline{f}_{nc}(x,u,t)=-f_{nc}(x,u,t)$.

Following Eq.~\eqref{ls} (factorized for the Markovian dynamics),
the infinitesimal entropy discharged into the surrounding medium reads
\begin{align} \label{first}
  &  \delta s_m(t)  = \ln \frac{\exp \{-[m du_t - d\alpha(x_t,u_t,t)]^2/(4\gamma T dt)\}}{\exp \{-[m du_t - d\alpha^*(x_{t+dt},-u_{t+dt},t+dt)]^2/(4 \gamma T dt)\}} = \\ \begin{split}
  & = -\frac{1}{4\gamma T dt} \left\{[m du_t + \gamma u_t dt + \phi'(x_t) dt - f_{nc}(x_t,u_t,t) dt]^2 \right. \\ & \;\;\;\; \left. - [m du_t -\gamma u_{t+dt} dt + \phi'(x_{t+dt}) dt - \overline{f}_{nc}(x_{t+dt},-u_{t+dt},t+dt)dt]^2 \right\} = \end{split} \\ 
\begin{split}
  &= -\frac{1}{\gamma T} \left[ m \gamma du_t \circ u_t + \gamma u_t \phi'(x_t) dt - \gamma \frac{u_t f_{nc}(x_t,u_t,t) + u_{t+dt} \overline{f}_{nc}(x_{t+dt},-u_{t+dt},t+dt)}{2}dt \right. \\
  &\;\;\;\;  - m du \frac{f_{nc}(x_t,u_t,t)- \overline{f}_{nc}(x_{t+dt},-u_{t+dt},t+dt)}{2}  \\ 
 & \;\;\;\; \left. - \frac{\phi'(x_t) f_{nc}(x_t,u_t,t)-\phi'(x_{t+dt}) \overline{f}_{nc}(x_{t+dt},-u_{t+dt},t+dt)}{2} dt\right].  \label{last}
\end{split}
\end{align}
In the first passage we have used the fact that the time-reversal of
$du_t$ is $-u_{t}-(-u_{t+dt}) = du_t$. In the second passage we have
neglected terms which goes to zero faster than
$dt$ and we have replaced $du_t (u_t + u_{t+dt})/2$ with $du_t
\circ u_t$.

The cases considered in this paper are the following:

\begin{itemize}

\item The standard case in Eq.~\eqref{generic} where $f_{nc}(t)$ is even and external (i.e. it does not
depend upon $x,v$): $\overline{f}_{nc}(t)=f_{nc}(t)$. In this case the last two terms in Eq.~\eqref{last} become of higher order in $dt$
and one gets:
\begin{equation}
\delta s_m(t) = -\frac{1}{T} u_t \circ [m du_t + \phi'(x_t)dt - f_{nc}(t)dt]
\end{equation}
which immediately gives Eq.~\eqref{seifert}.

\item The case considered in~\cite{mandal}, where $f_{nc}(x,u,t)=-\tau
  u \phi''(x)$ and (``conjugated entropy production'') $\overline{f}_{nc}(x,u,t)=\tau u \phi''(x)$. In
  this case the last term becomes of higher order in $dt$, while the term $du^2$ cannot be discarded (as it contains $dW^2 \sim dt$), and therefore one gets
\begin{equation}
\delta s_m(t) = -\frac{1}{T}\left\{u_t \circ [m du_t + \phi'(x_t)dt - f_{nc}(x_t,v_t,t) dt] + \frac{\tau^2}{2}du_t^2 \phi''(x_t)\right\},
\end{equation}
(where we have used $m \equiv \mu = \gamma \tau$), that is Eq.~\eqref{mand}.

\item The case considered in~\cite{BPM17} and in~\cite{FNCTVW16},
  where $f_{nc}(x,u,t)=-\tau u \phi''(x)$ and (according to the
  standard definition of stochastic entropy production) $\overline{f}_{nc}(x,u,t)=-\tau u \phi''(x)$. In
  this case the third term in Eq.~\eqref{last} is of higher order   in $dt$. All the other terms must be kept, giving
\begin{equation} \label{lsok}
\delta s_m(t) = -\frac{1}{T} \left(1+\frac{\tau}{\gamma} \phi''(x_t)\right) u_t \circ [m du_t + \phi'(x_t)dt].
\end{equation}
If no terms are neglected, it gives exactly Eq.~\eqref{newcla}.

\item If in Eq.~\eqref{lsok} the exact differentials ($u_t \circ du_t
  = d u_t^2 /2$, $u_t \phi'(x_t) dt = d\phi(x_t)$ and $u_t
  \phi'(x_t)\phi''(x_t) dt= d[\phi'(x_t)]^2/2$) are removed, then only one
  terms remains:
\begin{equation} \label{lsappr}
\delta s_m(t) \approx -\frac{1}{T} \tau^2 \phi''(x_t) u_t \circ du_t,
\end{equation}
ie. Eq.~\eqref{fred}.

\item If a time-dependent potential is considered, then a second
  non-conservative force appears $f_{nc,2}(x,t)=-
  \tau\partial_x\partial_t \phi(x,t)$. We stress that the dependence
  upon time of $\phi(x,t)$ is external, i.e. (keeping the standard
  recipe of stochastic thermodynamics) the probability of the reversed
  dynamics is generated by the same equation, that is no change of
  sign is attributed to $\partial_t$. Basically we have
  $\overline{f}_{nc,2}(x,t)=- \tau\partial_x\partial_t
  \phi(x,t)$. Introducing $f_{nc,2}$ in Eq.~\eqref{first} leads to the
  appearance of two new addends in the brackets $[...]$ of
  Eq.~\eqref{last}: one totally new term $-\tau u_t \partial^2
  \phi(x_t,t) f_{nc,2}(x_t,t)$ coming from the product $f_{nc}
    f_{nc,2}$; one surviving term $-\gamma u_t f_{nc,2}(x_t,t)$ in
    the third addend. No new terms appear inside the fourth and fifth
    addend.  In conclusion one gets
\begin{equation} \label{lsok2}
\delta s_m(t) = -\frac{1}{T} \left(1+\frac{\tau}{\gamma} \partial_x^2\phi(x_t,t)\right) u_t \circ [m du_t + \phi'(x_t)dt - f_{nc,2}(x_t,t)],
\end{equation}
which gives again Eq.~\eqref{newcla}.

\end{itemize}

\bibliography{biblio}

\end{document}